# Long range beam-beam interactions in PEP-II

M. E. Biagini
*LNF, INFN, Frascati, Italy*

The PEP-II luminosity upgrade foreseen for the next years requires an increased number of bunches and lower $\beta_y$* with minor modifications to the present Interaction Region (IR2). When increasing the collision frequency the beams separation in IR2 can be an issue. A study of the effect of the parasitic crossings for both the head-on and horizontal crossing angle options is presented.

## 1. INTRODUCTION

An upgrade of the PEP-II B-Factory luminosity to values $\geq 3 \times 10^{34}$ $cm^{-2}$ $s^{-1}$ is presently under study [1]. This increase in luminosity requires higher currents, larger number of bunches and lower values of $\beta_y$*. This last option requires in turn that the bunch length $\sigma_z$ is reduced accordingly to avoid as much as possible the hourglass effect. The option of introducing a small horizontal crossing angle in order to minimize the effect of the parasitic collisions is presently under study. A parametric study of the effect of long range beam-beam interactions is presented in this paper.

## 2. PEP-II LUMINOSITY UPGRADE

The possibility to work with a bunch pattern filling every other bucket (called by_2) is being explored at PEP-II. An estimate of the parasitic crossings (PC) effect on the linear beam-beam parameter can be very useful to evaluate if the introduction of a small crossing angle can help to decrease the PC effect on the beam-beam parameters.

Two options are presently under study [2]:

a) head-on collision with improved vertical focusing, for lower $\beta_y$*, provided by additional permanent magnet (pm) material between B1 (the small pm bending magnet used to separate the beams outside the IP) and Q1 (the first vertical focusing quadrupole);
b) small horizontal crossing angle collision, used to increase the number of colliding bunches with lower impact from PC. The improved vertical focusing could be provided in this case by substituting 5 B1 slices with 5 pm quadrupole slices.

A preliminary study of the latter option has shown that with the present IR2 layout the corrector strengths are able to cope with a crossing angle ranging from 0. to ±3.5 mrad, leaving the orbits outside IR2 unperturbed.

In order to be able to choose the new IR2 configuration it is necessary:

1) to evaluate the luminosity and beam-beam linear tune shifts for different values of $\beta_y$* and $\sigma_z$, by taking into account the hourglass effect [3,4], to have an evaluation on how much the luminosity would be reduced if the bunch length could not be shortened enough;
2) to compute the PC tune shifts as a function of the horizontal half crossing angle $\theta$, in order to be able to choose the smallest angle value providing enough separation and smaller PC tune shifts.

In Table 1 the parameters of the two beams used for this evaluation are listed. For sake of simplicity these values have been kept constant in all calculations, while $\beta_y$* and $\sigma_z$ have been used as free parameters.

Table 1: Beam parameters for LER and HER

| | **LER** | **HER** |
|---|---|---|
| **I (mA)** | 4500 | 2000 |
| **N. bunches** | 1700 | 1700 |
| **Npart/bunch** | $1.22 \times 10^{11}$ | $5.4 \times 10^{10}$ |
| **$\beta_x$* (cm)** | 25 | 25 |
| **$\varepsilon_x$ (nm)** | 40 | 40 |
| **$\varepsilon_y$ (nm)** | 1.2 | 1.1 |

## 3. CROSSING ANGLE

The crossing angle geometry has many advantages: it allows for a higher collision frequency, so that a larger number of bunches can collide, the beams are "naturally" separated as soon as they leave the collision point, so there is no need for dipoles close to the IP, and the beams can be sooner accommodated in two separate rings. These are the reasons why "factories", as DAΦNE, CESR and KEK-B, have chosen it.

However the crossing angle geometry has also some drawbacks.

- luminosity and tune shifts are "geometrically" reduced, as will be discussed in section 5;
- larger vacuum chamber aperture is needed;
- the beams travel off-axis in the quadrupoles, where field quality is degraded. Non-linear fields and fringing field effects have then to be carefully taken into account when modelling the beam trajectory;
- with a large crossing angle, highly desirable from a "geometric" point of view, synchro-betatron resonances, which couple the transverse and longitudinal phase space, can be excited with a resulting increase of the beam spot size at the IP and a consequently lower luminosity.

The Piwinski angle, defined as:

$$\Theta = \theta\, \sigma_z/\sigma_x$$

where θ is the half crossing angle and $\sigma_x$ and $\sigma_z$ are the horizontal and longitudinal beam sizes, is a parameter used to estimate how dangerous the crossing angle can be. Up to now DAΦNE and KEK-B are the storage rings where θ has reached higher values with some loss in luminosity due to beam blow up but no destructive effects; however this parameter should in general be kept as low as possible, and it could be a limitation when trying to reach very high beam-beam tune shift values. For a comparison, in Table 2 an evaluation of the Piwinski angle for the Factories working with a crossing angle is presented. In the last column the Piwinski angle for PEP-II is computed for $\beta_x^* = 25$ cm and $\varepsilon_x = 40$ nm.

Table 2: Crossing angle

| | CESR | DAΦNE | KEK-B | PEP-II |
|---|---|---|---|---|
| $\sigma_x^*$ (μ) | 470 | 1440 → 1010 | 103 | 100 |
| $\sigma_z$ (cm) | 1.8 | 2 | 0.54 | 0.5 |
| θ (mrad) | ±2.3 | ±12 → ±14.5 | ±11 | ±3.5 |
| Θ (mrad) | 0.09 | 0.17 → 0.29 | 0.57 | 0.18 |

In conclusion, we think that the choice to collide with or without a crossing angle must be a trade-off between the aforementioned effects and the PC effect, described in the following section. It is important to determine the minimum beam separation required in order to have acceptable beam-beam tune shifts at the PC and reasonable lifetimes: this sets the choice on the θ value.

## 4. PARASITIC CROSSINGS

When the bunch spacing is reduced the beams travel in the same pipe with a smaller separation and can interact with destructive effects at the PC: a distance between beam cores of at least 10 $\sigma_x$ is required at the first PC (the most harmful) in order not to have the beam tails seeing each other, with a consequent decrease in lifetime.

Moreover the long range beam-beam interactions can become as important as the IP one and the luminosity is degraded. This reduction can be estimated only by a beam-beam simulation including the PC effect, while the PC tune shifts can be computed once we know the beams separation at the PC. For the by_2 pattern the first PC in PEP-II is located at 0.63 m from the IP. Each bunch experiences this crossing twice, coming to and from the IP.

The tune shifts due to the PCs can be estimated, for Gaussian beams, by the following formulae [5]:

$$\xi_x = -\frac{N r_e}{2\pi\gamma}\frac{\beta_x\left(x^2-y^2\right)}{\left(x^2+y^2\right)^2}$$

$$\xi_y = +\frac{N r_e}{2\pi\gamma}\frac{\beta_y\left(x^2-y^2\right)}{\left(x^2+y^2\right)^2}$$

where x and y are the horizontal and vertical beam separation, N is the number of particles in the opposite bunch, γ is the beam energy. As it is shown by these formulae it is just the absolute value of the separation that counts from the beam-beam point of view, and not the number of $\sigma_x$, which limits the lifetime instead. In the following calculations the absolute values of $\xi_x$ and $\xi_y$ at the PC have been taken.

## 5. LUMINOSITY AND TUNE SHIFTS

Unfortunately there is not only the PC tune shift issue to limit the collider's performances. With a crossing angle we can get rid of the previous problem, with some costs (as the pipe aperture in the IR), but another issue arises: luminosity and horizontal beam size degrade when introducing the crossing angle.

The geometric effect of a horizontal crossing angle θ on luminosity and beam-beam tune shifts needs to be studied with 3D beam-beam simulations, however to give a first estimate it can be computed following Refs. [6,7]. For the case when γ >> tg(θ/2) the luminosity and tune shifts formulae are simply:

$$L = \frac{N^2}{4\pi\sigma_y\sqrt{\left(\sigma_z^2 tg^2(\theta/2)+\sigma_x^2\right)}}$$

$$\xi_{xp} = \frac{r_e N}{2\pi\gamma}\frac{\beta_x}{\sqrt{\left(\sigma_z^2 tg^2(\theta/2)+\sigma_x^2\right)}\left(\sqrt{\left(\sigma_z^2 tg^2(\theta/2)+\sigma_x^2\right)}+\sigma_y\right)}$$

$$\xi_{yp} = \frac{r_e N}{2\pi\gamma}\frac{\beta_y}{\sigma_y\left(\sqrt{\left(\sigma_z^2 tg^2(\theta/2)+\sigma_x^2\right)}+\sigma_y\right)}$$

with the usual meaning of the symbols.

These formulae are derived from the formulae for head-on collision, by just substituting the horizontal beam size by:

$$\left(\sigma_x^2+\sigma_z^2 tg^2(\theta/2)\right)^{1/2}$$

Then, the effect of the crossing angle is to increase the effective horizontal beam size by a factor $\sigma_z$ tg (θ/2). Therefore luminosity and the tune shifts are reduced, with the horizontal tune shift dropping faster than luminosity and vertical tune shift.

As an example, in Figs. 1 and 2 the luminosity as a function of the crossing angle, for different values of the $\beta_y^*$ and for two values of the bunch length ($\sigma_z$ = 9 and 7 mm) is plotted. The hourglass reduction factor has been also taken into account [4]. A very small crossing angle, smaller than ±2 mrad, has practically no effect on the luminosity, while the design luminosity could in principle be reached also with a longer bunch length (9 mm). The geometric effect on both tune shifts is shown in Figs. 3 and 4 for LER and HER, for a bunch length $\sigma_z$ = 9 mm, hourglass aggravating factors included.

However strong-strong beam-beam simulations (Cai [8], Ohmi) taking into account the crossing angle show for PEP-II parameters a more severe decrease in luminosity and maximum achievable tune shift at the main IP. Further studies are needed with different sets of parameter to check this point.

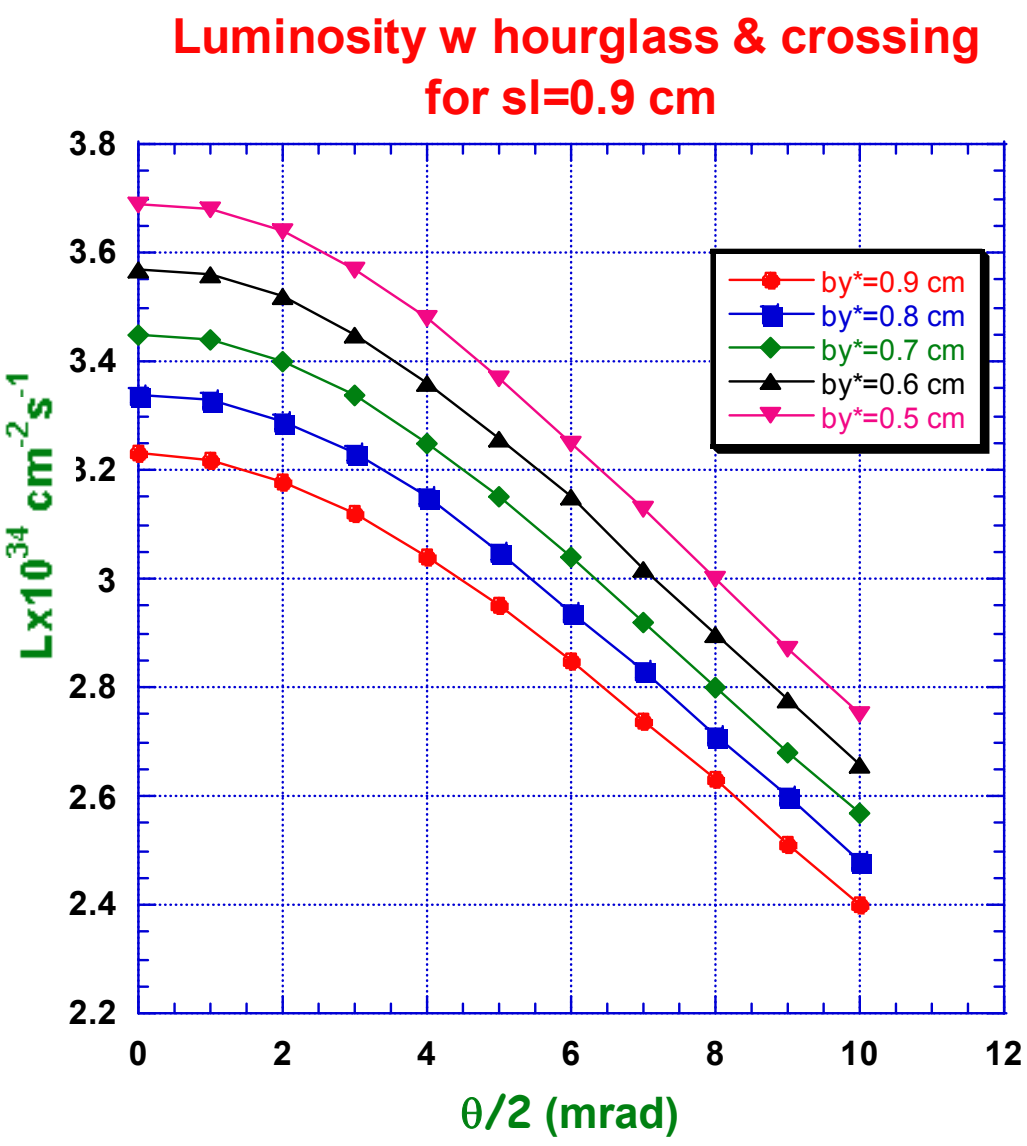

Fig. 1 – Luminosity vs θ/2 for different $\beta_y$*, $\sigma_z$ = 9 mm.

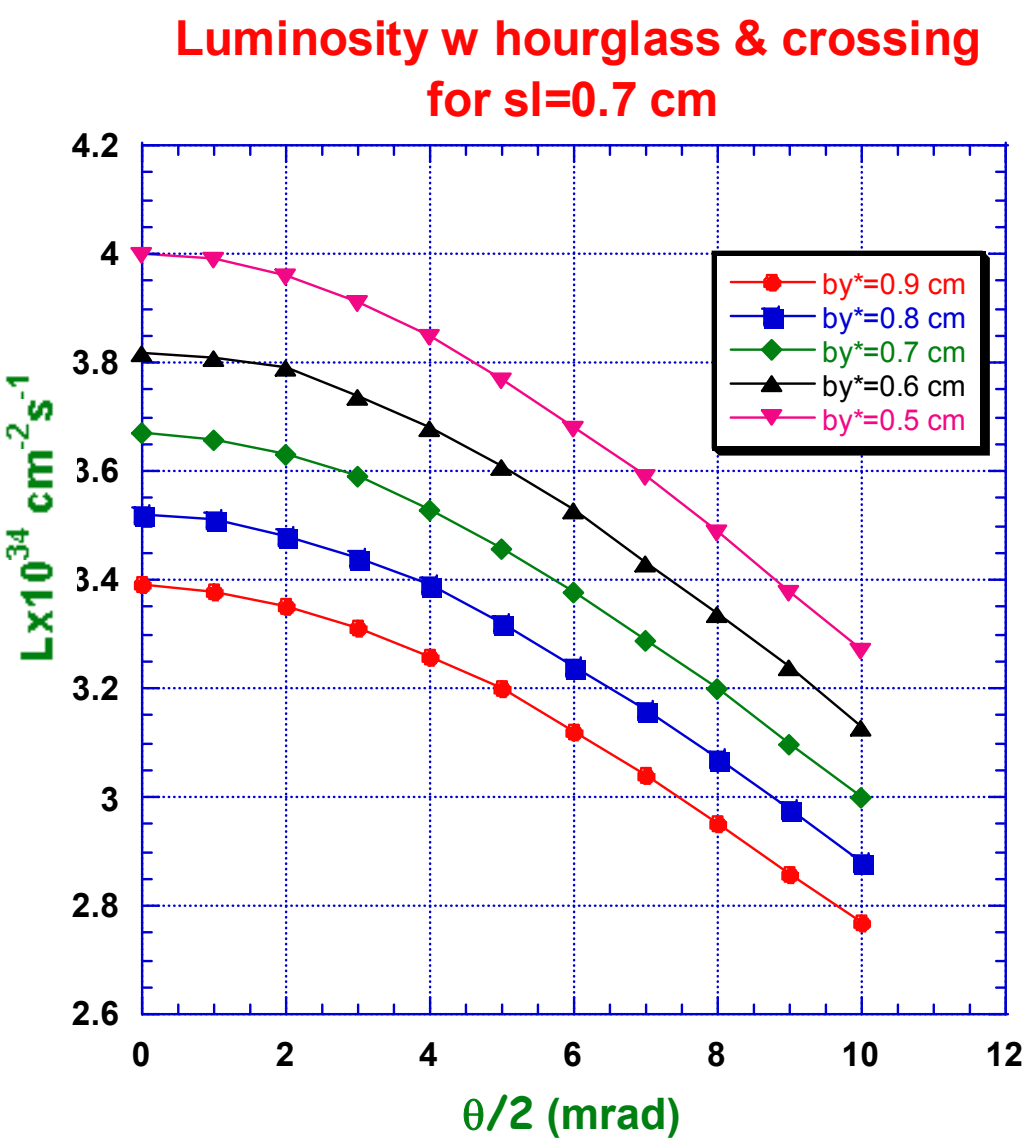

Fig. 2 – Luminosity vs θ/2 for different $\beta_y$*, $\sigma_z$= 7 mm.

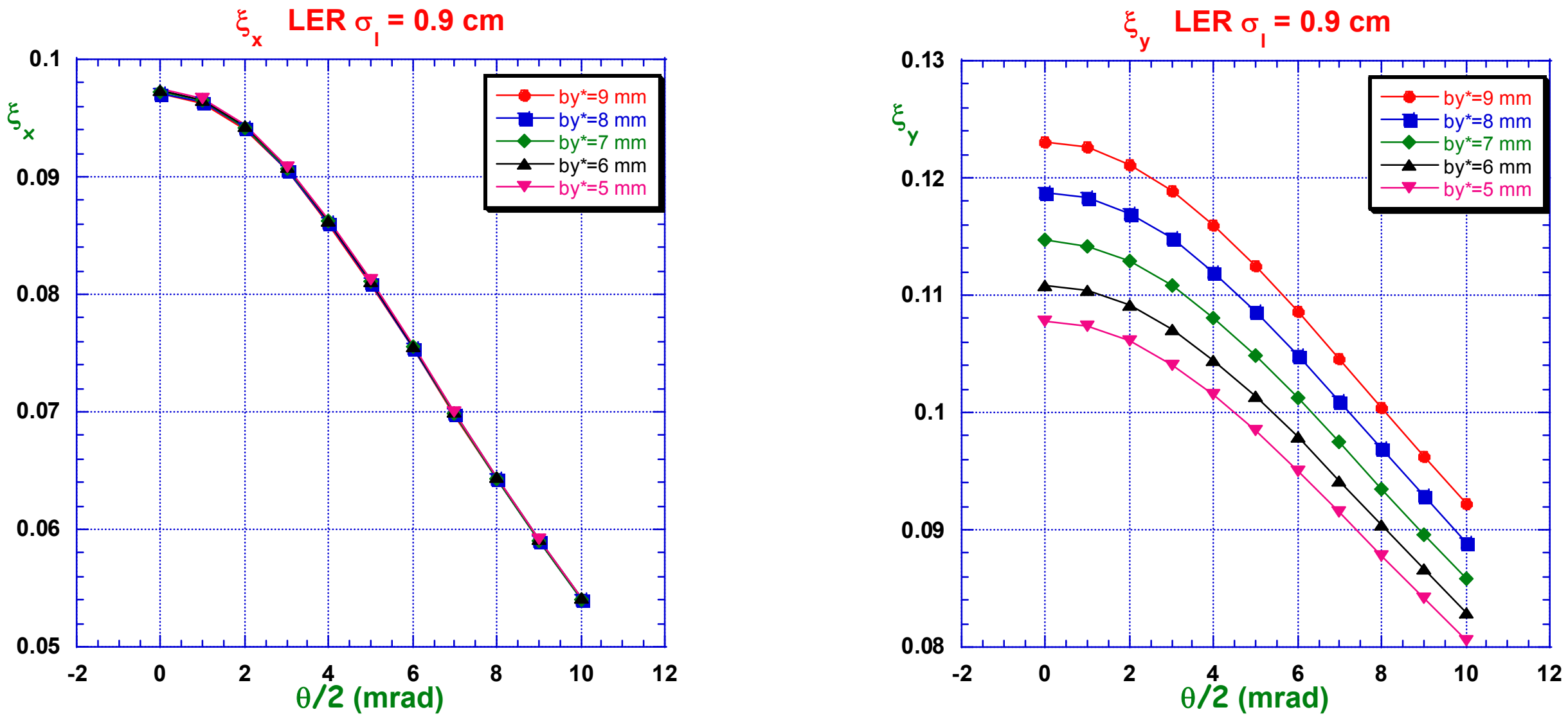

Fig. 3 – LER IP tune shifts (left: horizontal, right: vertical) vs θ/2 for different $\beta_y$* and $\sigma_z$ = 9 mm.

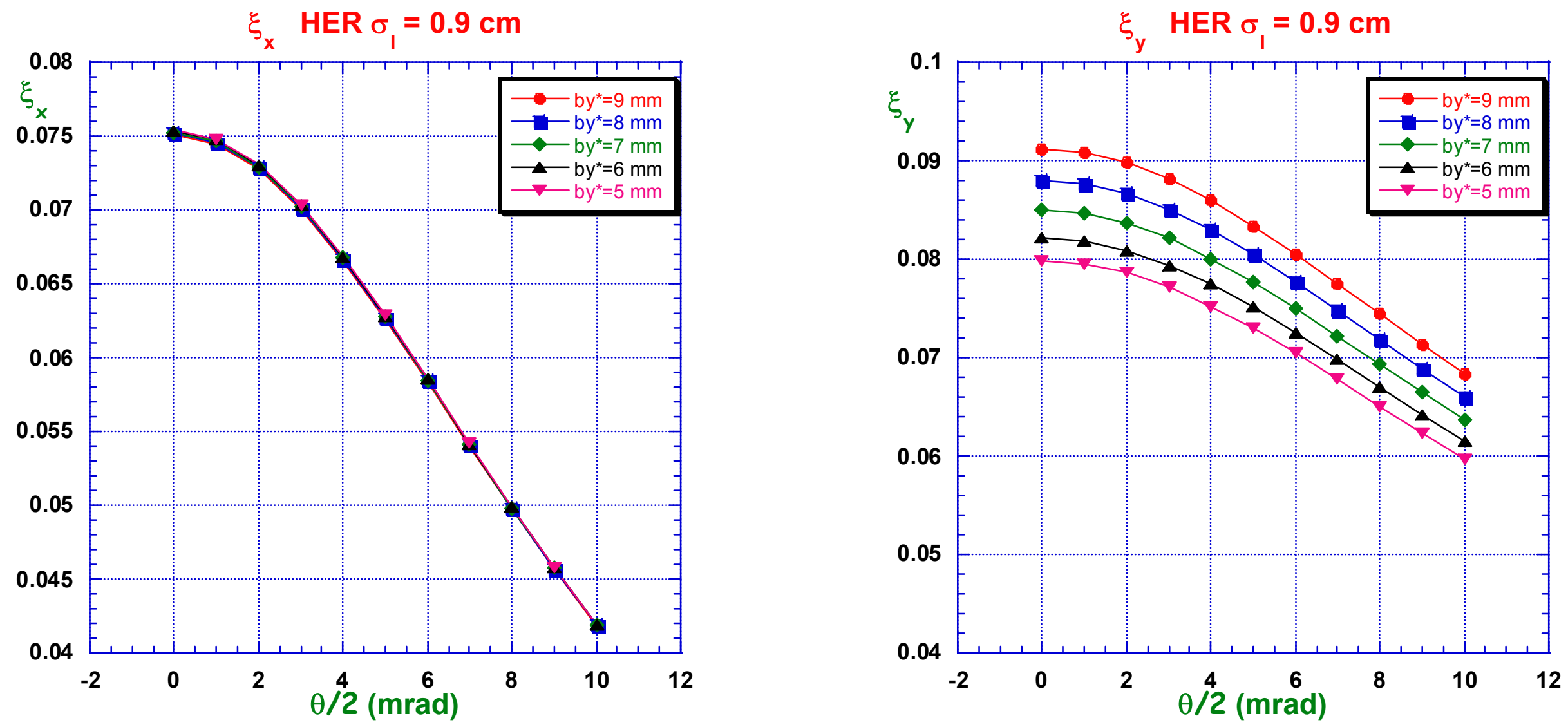

Fig. 4 – HER IP tune shifts (left: horizontal, right: vertical) vs θ/2 for different $\beta_y$* and $\sigma_z$ = 9 mm.

## 6. WORKING IN A BY_2 PATTERN

The PEP-II luminosity upgrade is designed with 1700 bunches, that is a by_2 bunch pattern. We concentrate our analysis on the 1st PC, which is clearly the most harmful.

The PC tune shifts have been computed as a function of the $\beta_y$* for different IR geometry, from head-on collision to ±10 mrad crossing angle. In Figs. 5 and 6 the LER and HER 1st PC tune shifts are plotted, with $\beta_y$* ranging from 5 to 9 mm and hourglass factor included.

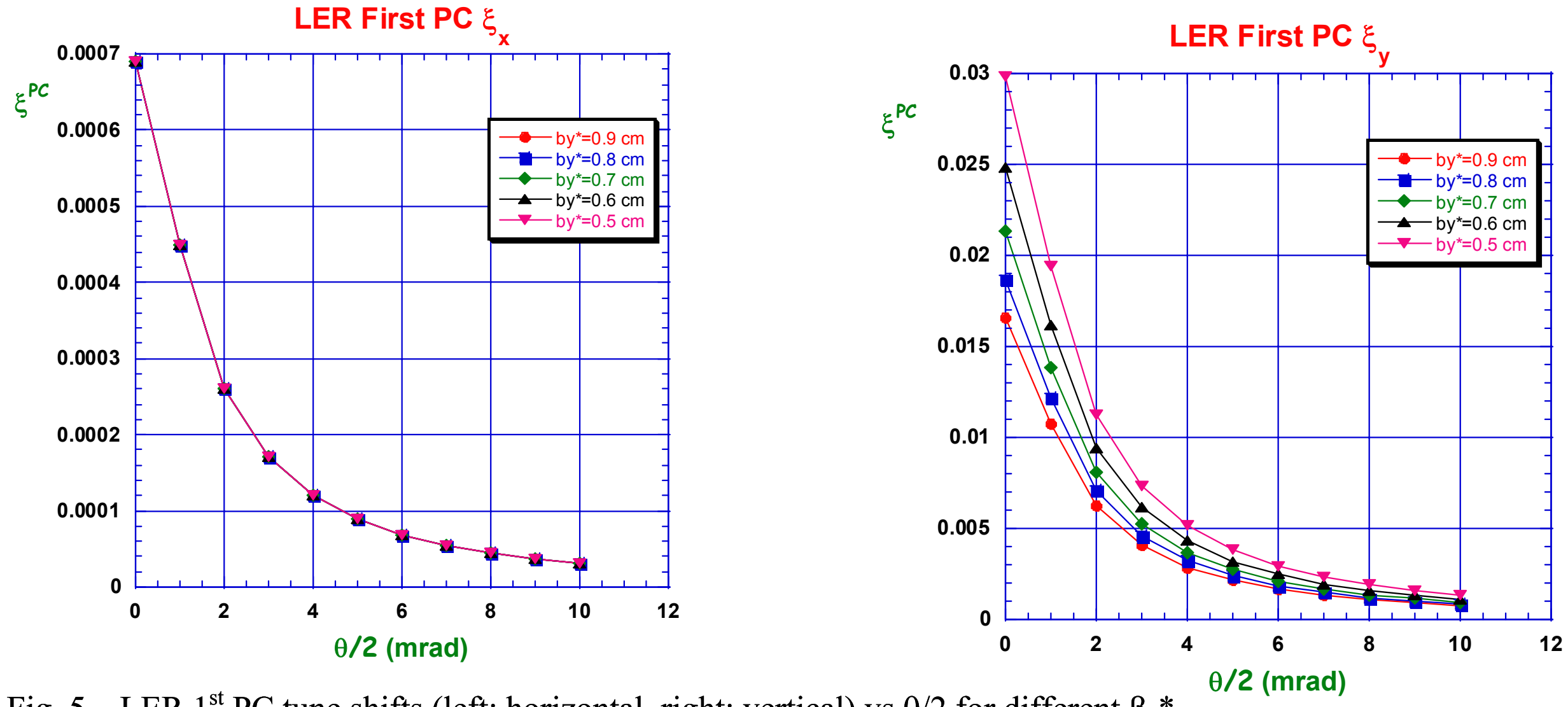

Fig. 5 – LER 1st PC tune shifts (left: horizontal, right: vertical) vs θ/2 for different $\beta_y$*.

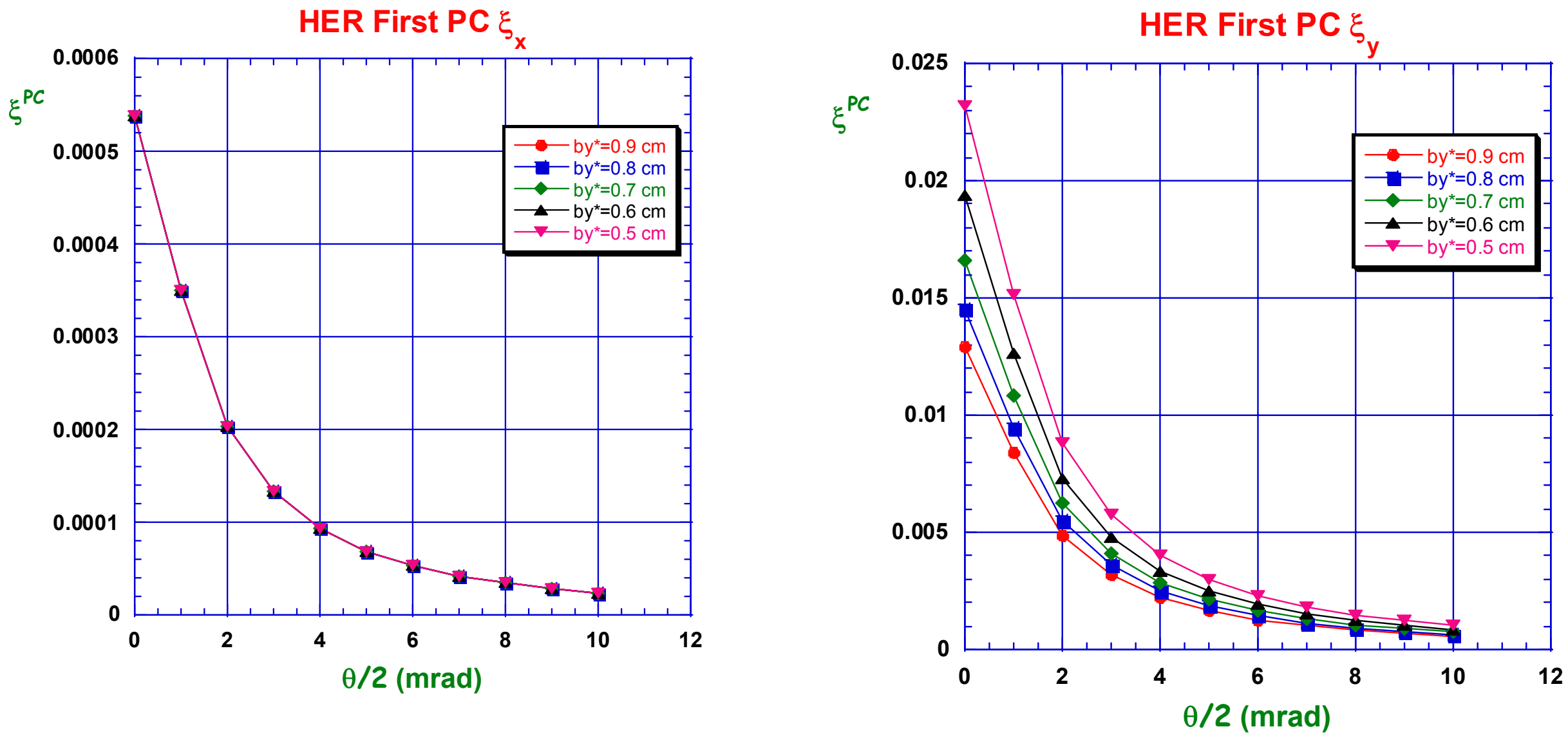

Fig. 6 – HER 1st PC tune shifts (left: horizontal, right: vertical) vs θ/2 for different $\beta_y^*$

The horizontal tune shift is clearly negligible already at the 1st PC, while the vertical one can reach remarkably high values. Of course increasing the value of the crossing angle the tune shifts rapidly decrease.

To estimate more clearly how the PC tune shifts can aggravate the beam-beam interaction, their value for the 1st PC was also compared to the main IP tune shift. For this purpose the main IP tune shifts were scaled by decreasing the bunch length accordingly so to keep the design luminosity constant ($3.3 \times 10^{34}$ cm$^{-2}$ s$^{-1}$), including the hourglass aggravating factors.

As an example, in Figs. 7 and 8 the absolute value of the PC $\xi_y$, normalized to the IP ones, is plotted for two values of the bunch length (9 and 7 mm) and for different $\beta_y$* values. Note that the PC tune shift contribution has to be counted twice since each bunch experiences a PC collision on both sides of the IP. The impact of the PC collision is of course stronger for lower $\beta_y$* and shorter bunches, going up to 60% for $\beta_y$* = 5 mm, $\sigma_z$ = 7 mm in head-on collision. In this case however even a small ±1 mrad crossing angle could reduce the effect from 60% to 40%.

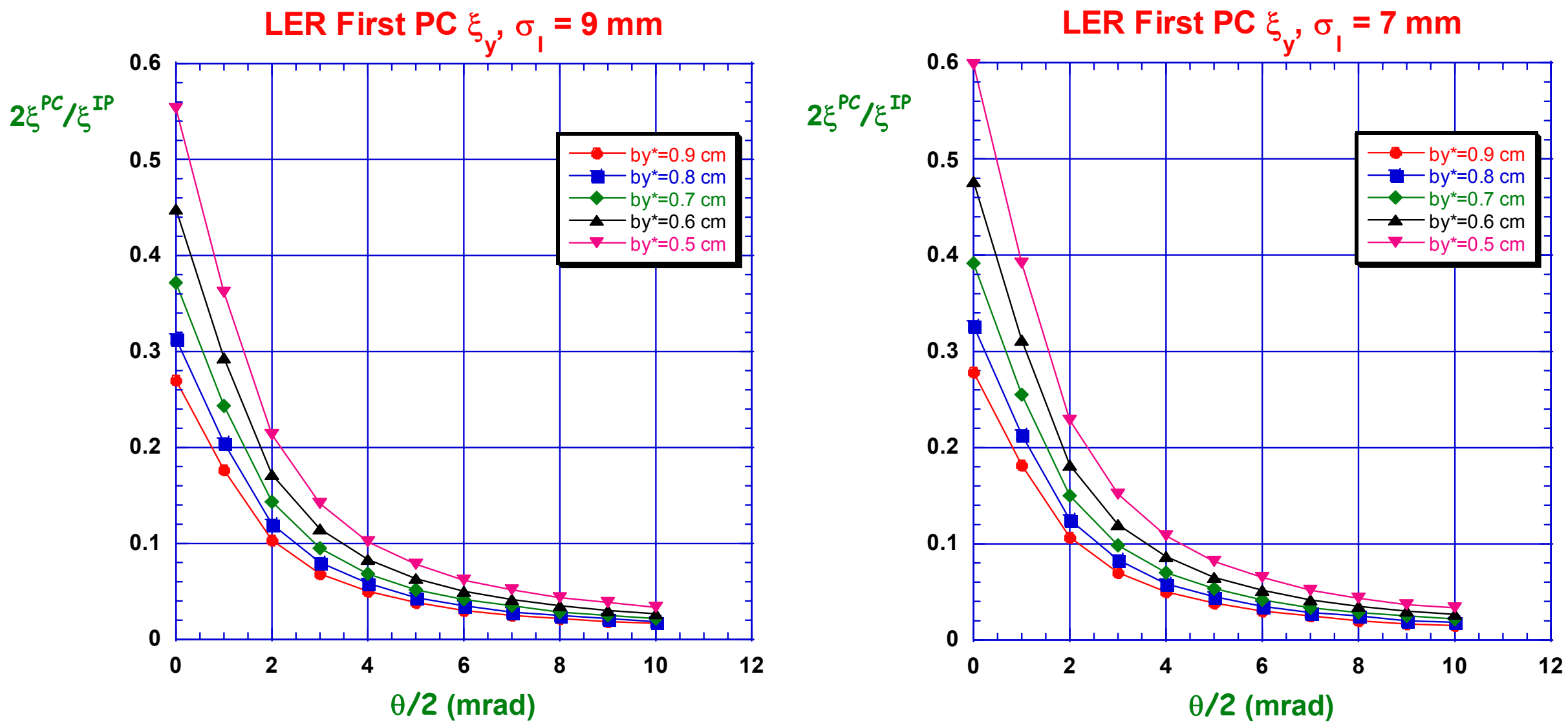

Fig. 7 – LER 1st PC $\xi_y$, normalized to the main IP one, vs θ/2 for different $\beta_y$* (left: $\sigma_z$ = 9 mm, right: $\sigma_z$ = 7 mm).

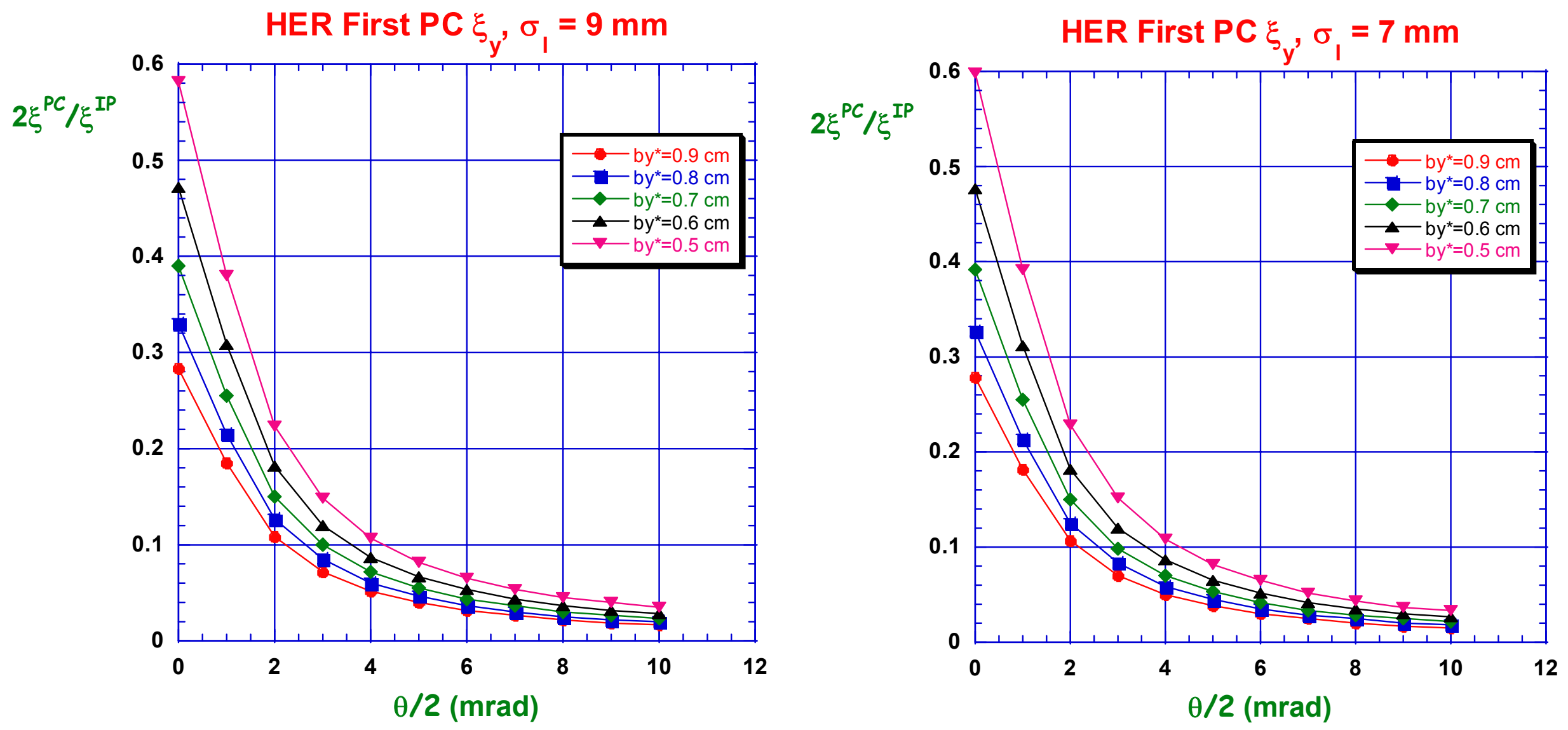

Fig. 8 – HER 1st PC $\xi_y$, normalized to the main IP one, vs $\theta/2$ for different $\beta_y^*$ (left: $\sigma_z = 9$ mm, right: $\sigma_z = 7$ mm).

## 7. REMARKS AND CONCLUSIONS

The introduction of a crossing angle, to increase the number of colliding bunches, can reduce the strength of the PC collisions but also reduces the maximum achievable luminosity.

The PC tune shifts rapidly decrease as a function of the crossing angle. However it seems that the introduction of a very small crossing angle (≤2 mrad) could be a safe option which does not affects too much the luminosity. In fact it can decrease the PC effect to less than 10%, still probably tunable by the operators during delivery (as remarked by F.J. Decker). Such a value of the crossing angle is rather easy to introduce in the present PEP-II IR design and manageable from the point of view of the present machine correctors. As a test, an IR design with tunable crossing angle from 0 to 2 mrad could be implemented for the future operation with a by_2 pattern at high currents. The head-on collision can still be an option if the actual luminosity turns out to be strongly affected by such a small crossing angle.

Few questions are still open: could it be possible to work with a smaller number of bunches with higher current per bunch to get the same peak luminosity, with the same tune shifts (as suggested by M. Placidi)? Or is it wiser to accept a degraded luminosity by the crossing angle but operate with a larger number of bunches and larger total beam current?

The key point however is represented by the bunch length: it has to be carefully studied if very short bunches (6. to 5. mm) can be obtained with minor modifications to the present PEP-II lattices. In any case analytically it seems that even with a longer bunch length, of the order of 9 to 7 mm, it could be possible to get the design luminosity.

Of course beam-beam simulations are the only way we have now to answer to these questions. It is mandatory to include the PC's and the crossing angle in a 3D strong-strong beam-beam simulation.

## Acknowledgments

The author wishes to thank J. Seeman and M. Sullivan for their encouragement, their support and for their valuable suggestions and discussions on this subject. The author is also indebted with M. Furman who kindly provided his code for the evaluation of the luminosity and tune shifts aggravating factors due to the hourglass effect.

## References

[1] J. T. Seeman, "Future very high luminosity options for PEP-II", this Workshop.

[2] M. K. Sullivan, "Upgrades to the PEP-II Interaction region", this Workshop.

[3] M. Furman, PEP-II ABC-21/ESG-technote-161, April 1991 (revised 10/03).

[4] M. Furman, program HG_csi.f, private communication.

[5] J. Jowett, "Beam-beam tune shifts for Gaussian beams", Handbook of Accelerator Physics and Engineering, World Scientific.

[6] O. Napoly, Particle Accelerators, 1993, Vol. 40, pp. 181-203.

[7] P. Raimondi, M. Zobov, "Tune shift in beam-beam collisions with a crossing angle", DAΦNE Tech.Note G-58, April 2003.

[8] Y. Cai, "Beam-Beam Simulation for PEP-II", this Workshop.